\documentclass[12pt]{article}
\usepackage[normalem]{ulem}
\usepackage[unicode=true]{hyperref}
\usepackage[utf8]{inputenc}
\usepackage{amsmath,bm,amssymb,bbm}
\usepackage{graphicx}
\usepackage{color}
\usepackage{geometry}
\usepackage{float}

\usepackage{wrapfig}
\usepackage[dvipsnames]{xcolor}
\usepackage[makeroom]{cancel}

\usepackage{tikz}
\usepackage{rotating}

\usepackage{pgfplots}
\usepackage{pgfplotstable}
\usepackage{tkz-fct}
\usetikzlibrary{calc,patterns}
\usetikzlibrary{math}
\usepackage{subfigure}
\usepackage{tkz-euclide}
\usetikzlibrary{decorations.pathmorphing}
\usetikzlibrary{arrows}
\usetikzlibrary{arrows.meta}
\usetikzlibrary{decorations.markings}

\newcommand{\itl}{\int\limits}

\definecolor{bostonuniversityred}{rgb}{0.8, 0.0, 0.0}
\definecolor{dukeblue}{rgb}{0.0, 0.0, 0.61}
\definecolor{ao(english)}{rgb}{0.0, 0.5, 0.0}
\definecolor{darkmagenta}{rgb}{0.55, 0.0, 0.55}
\definecolor{armygreen}{rgb}{0.29, 0.33, 0.13}
\definecolor{coquelicot}{rgb}{1.0, 0.22, 0.0}
\definecolor{fucsiak}{rgb}{0.4, 0.08, 0.4}
\definecolor{airforceblue}{rgb}{0.36, 0.54, 0.66}
\definecolor{brown(traditional)}{rgb}{0.59, 0.29, 0.0}
\definecolor{electricultramarine}{rgb}{0.25, 0.0, 1.0}
\definecolor{champagne}{rgb}{0.97, 0.91, 0.81}
\definecolor{chromeyellow}{rgb}{1.0, 0.65, 0.0}
\definecolor{cornellred}{rgb}{0.7, 0.11, 0.11}

\title{Alpha-effect in three-dimensional vortex of conducting rotating liquid}
\author{L.L. Ogorodnikov and S.S. Vergeles}

\begin{document}

\maketitle

\abstract{We study one-point statistics of spiral turbulent pulsations on the background of three-dimensional large-scale vortex in a rotating fluid. Assuming that the helical flow is created by a statistically axially symmetric random force with broken mirror symmetry, we analytically calculate velocity-vorticity mean including its magnitude and the anisotropy. For electrically conducting liquid, we examine the $\alpha$-effect in the system. The found elements of $\alpha$-matrix strongly depend on relation between Rossby $\mathrm{Ro}$ and magnetic Prandtl $\mathrm{Pr}_{\mathrm{m}}$ numbers in considered region $\mathrm{Ro}\lesssim 1, \,\ \mathrm{Pr}_{\mathrm{m}} \lesssim 1$. We establish criterion for the numbers when the $\alpha$-effect leads to instability of large-scale magnetic field.

}

\section{Introduction}

Flow of a conducting liquid is inevitably accompanied by electric current, if the flow has sufficient amplitude \cite{rudiger2006magnetic}. It is convenient to describe the current in terms of the induced magnetic field. The magnetic field can be considered as a result of amplification of some weak seminal magnetic field by the flow with further saturation of the amplification due to the back reaction of the magnetic field onto the flow \cite{moffatt1978field}, \cite{rincon2019dynamo}, \cite{tobias2021turbulent}. The examples of saturated statistically steady state are planetary magnetic fields \cite{jones2011planetary}, in particular Earth magnetism \cite{schaeffer2017turbulent,zeldovich1990almighty}, coherent magnetic field in the Universe \cite{subramanian2019primordial} and a model isotropic turbulent state \cite{schekochihin2004simulations}. It is often that a large-scale magnetic field $\overline{\bm B}$ is maintained on the background of small-scale turbulent motion. In this case, the amplification of the magnetic field by the flow is reasonable to describe by coefficients which are the result of averaging over fast and small-scale turbulent pulsations of the flow \cite{rudiger2006magnetic}, \cite{brandenburg2005astrophysical}. The interaction of the turbulent pulsations with the large-scale magnetic field produces a small-scale weak magnetic field component, which in turn interacts with the turbulent flow and produces large-scale electromotive force ${\bm{\mathcal{E}}}$. If the turbulent flow is characterized by nonzero mean helicity, the electromotive force is proportional to the magnetic field, ${\bm{\mathcal{E}}} = \hat \alpha \overline{\bm B}$. Pseudo-tensor $\hat \alpha$ is of the second rank and the proportionality is called $\alpha$-effect.

There are several mechanisms in the process of dynamo formation \cite{tobias2021turbulent, sokoloff2014dynamos}. Here we consider the liquid rotating fast as a whole with angular velocity $\Omega$.  For example, this is the case of liquid core in Earth \cite{olton2015coredynamics} and other planets \cite{jones2011planetary}, whereas the cosmological magnetic dynamo occurs in other conditions and therefore is not considered here. We assume here that the flow is substantially three-dimensional, so the small-scale turbulent pulsations are inertial waves \cite{dormy2007mathematical}. The fast rotation suppresses the direct three-dimensional turbulent cascade \cite{proudman1916motion} and supports the inverse cascade which can turn into a coherent vortices under certain conditions \cite{kolokolov2020structure}, \cite{seshasayanan2018condensates}, \cite{guzman2022flow}. We assume that such a vortex is established, so there is a mean large-scale geostrophic vortex flow ${\bm U}$. An axially symmetric flow by itself does not cause an exponential growth of magnetic field in time \cite{moffatt1978field, tobias2021turbulent, cowling1933magnetic}. The violation of such symmetry, in particular, due to the small-scale turbulent pulsations \cite{braginskii1964theory,braginsky1990some}, may lead to the magnetic field growth. In order for the small-scale pulsations to ensure the growth of the magnetic field via the $\alpha$-effect, its statistics should not be invariant under the mirror symmetry \cite{rudiger2006magnetic}. In general, the violated parity-invariance means that the turbulent pulsations possess mean helicity, although the nonzero mean helicity is not necessary property of such violation \cite{gilbert1988helicity}. Violation of spatial parity by turbulent flow statistics can be achieved due to the inhomogeneity of the driving force, as it is the case in geodynamo \cite{davidson2014dynamics,ranjan2018internally}, or due to the superposition of convective flow and rotation as it is for solar dynamo in each hemisphere \cite{stix1976differential, charbonneau2023evolution}. In experiments this parity can be violated artificially \cite{gailitis2008history}.

We assume that the following inequalities are fulfilled for dimensionless parameters of the problem in the analytical calculation of tensors $\hat\alpha$ and $H_{ij}=\langle u_i\omega_j\rangle$, where ${\bm u}$ is the velocity field of turbulent pulsations, vorticity ${\boldsymbol\omega} = \mathop{\mathrm{curl}}{\bm u}$, angle brackets mean averaging over statistics of turbulent pulsations. Magnetic Prandtl number $\mathrm{Pr}_{\mathrm{m}} = \nu/\kappa$ is assumed to be small, $\mathrm{Pr}_{\mathrm{m}} \ll 1$, where $\nu$ is kinematic viscosity and $\kappa$ is magnetic diffusion coefficient in liquid. This inequality is fulfilled in the cases of planetary dynamo \cite{roberts2000geodynamo}, \cite{king2010convective} and laboratory dynamos on liquid metals \cite{sisan2004experimental}, \cite{figueroa2013modes}, \cite{gailitis2000detection}, \cite{stieglitz2001experimental}. The flow in a large-scale vortex is relatively weak compared to the rotation of the liquid as a whole, that is, the Rossby number is small $\mathrm{Ro}= |\Sigma|/2\Omega\ll1$, where the local shear rate in the differential rotation of the vortex is $\Sigma = \rho\partial_\rho (U/\rho)$, $U$ is azimuth component of the mean current, $\rho$ is distance to the vortex axis. Therefore the dynamics of turbulent pulsations in the main approximation is the dynamics of inertial waves. We assume that inertial waves are excited by a random force, statistically homogeneous in space and axisymmetric with respect to the direction of the rotation, but having a violation of mirror symmetry. The presence of both the turbulent pulsations and the large-scale differential rotation means that the $\alpha\Omega$-dynamo mechanism is realised \cite{parker1970origin, stix1974comments}. Further, we assume that the influence of the large-scale flow on the turbulent pulsations prevails over the viscous damping, $\nu k_f^2/|\Sigma|\ll1$, where $k_f$ is the characteristic wavenumber of the turbulent pulsations. The inequality can be rewritten as $\mathrm{Ek}\ll \mathrm{Ro}$, where the Ekman number is ${\mathrm{Ek}}=\nu k_f^2/2\Omega$. We consider arbitrary relation between the global rotation velocity $\Omega$ and the ohmic dissipation rate $\kappa k_f^2$ for a small-scale magnetic field component. We introduce the basic relations and calculate $\hat H$-tensor in Section~\ref{sec:start}.

In general, tensors $\hat H$ and $\hat \alpha$  are linearly independent, although both of them owe their origin to the violation of mirror symmetry in turbulence statistics. If the turbulence is statistically isotropic, then both of these tensors are proportional to the unit \cite{tobias2021turbulent}, and there is an estimation $\alpha \sim -\mathop{\mathrm{min}}(\tau_u,1/\kappa k_f^2)\cdot\hat H $ \cite{moffatt1970turbulent,krause2016mean,molchanov1985kinematic}. In the first limiting case, $\tau_u$ is the correlation time of the turbulent flow. If the fluid rotates rapidly as a whole and the statistics of the turbulent flow is axially symmetric, then tensors $\hat H$ and $\hat \alpha$ also inherit this symmetry. For transverse components there is an estimation $\alpha_{\scriptscriptstyle \perp} \sim -\mathop{\mathrm{min}}(1/\Omega,1/\kappa k_f^2)\cdot H_{\scriptscriptstyle \perp}$. The same relation is correct for the longitudinal component $\alpha_{\scriptscriptstyle \parallel}$, if the magnetic diffusion is strong: $\kappa k_f^2\gg\Omega$. In the inverse limit of weak magnetic diffusion, the longitudinal component is suppressed, $\alpha_{\scriptscriptstyle \parallel} \sim (\kappa k_f^2/\Omega)\alpha_{\scriptscriptstyle \perp}$ \cite{moffatt1970dynamo}. In \cite{brandenburg2002local} a numerical calculation of the magnetic dynamo was performed against the background of Kepler rotation, that has $\mathrm{Ro}=1/4$. The extracted $\alpha$-tensor has a strong asymmetry, so that $|\alpha_{\varphi\varphi}|\ll |\alpha_{\rho \rho}|$, where $\varphi$ is the azimuth angle that increases in the direction along the vortex streamlines. Our analytical calculations use the technique developed in the works \cite{kolokolov2020structure, ogorodnikov2022structure}. The results of our calculations presented in Section~\ref{sec:alpha} confirm the anisotropy of $\alpha$-tensor, obtained at \cite{brandenburg2002local}, although full compliance should not be expected due to the significant heterogeneity of the flow in the direction of the axis of rotation, assumed in \cite{brandenburg2002local}.

After calculating the $\hat\alpha$ tensor, we establish the criterion of dynamo instability of the axially symmetric large-scale magnetic field $\overline{\bm B}$ in the kinematic regime in Section~\ref{sec:mean-magnetic-field}, based on the general approach \cite{steenbeck1966berechnung}. According to \cite{braginsky1990some,brandenburg1997dependence}, in the conditions of a strong large-scale vortex flow, the growth of the magnetic field is determined by the element $\alpha_{\varphi\varphi}$. In \cite{brandenburg2002local} this criterion was obtained for magnetic field $\overline{\bm B}$, depending only on the coordinate along the axis of rotation $z$. We generalize it to the case of the dependence of $\overline{\bm B}$ also on the radial coordinate. Based on the obtained dependence of $\hat\alpha$ on the problems' parameters, we rewrite the criterion of the magnetic field instability in the form of a condition for dimensionless parameters characterizing the fluid and the flow. In Section ~\ref{sec:discussion} we discuss the applicability of our results to finite-size systems, and then check whether the obtained criterion is met in the known experimental MHD-setups. The calculations concerning $\alpha$-effect are collected in Appendix.

\section{Statistics of inertia waves in a coherent vortex under helical pumping}
\label{sec:start}

We start from the development of mathematical description for the small-scale turbulent pulsations excited by a helical small-scale driving force which is acting against background of a three-dimensional long-living columnar vortex that is in a rotating as a whole liquid. The aim of the Section is to trace how the velocity correlation function inherits the violation of the mirror symmetry from the helical force.

Similarly to \cite{kolokolov2020structure} and \cite{ogorodnikov2022structure}, we are considering turbulence in a rotating fluid with angular velocity ${\boldsymbol \Omega}$ and denote the direction of the rotation ${\bf e}_z = {\boldsymbol \Omega}/|{\boldsymbol \Omega}|$. In the rotating frame, the flow ${\bm v}(t,{\bm r})$ contains a mean component ${\bm U}$ and a fluctuating turbulent part ${\bm u}$, ${\bm v}={\bm U}+{\bm u}$. The mean component ${\bm U}$ is an axially symmetric vortex flow which changes at times and distances much larger than the characteristic ones of ${\bm u}$. The only nonzero component of the mean flow is the {azimuth} component $U$ in the cylindrical reference system $\{\rho,\varphi,z\}$ which axis coincides with the axis of the vortex. The global rotation is strong, so large-scale Rossby number is small,  $\mathrm{Ro}\sim |\Sigma|/2\Omega\ll 1$, where $\Sigma = \rho \partial_\rho \left( U / \rho\right)$ is the local large-scale shear rate. The temporal mean of the turbulent part of the flow is zero, $\langle {\bm u}\rangle=0$, and it is relatively weak, so its self-nonlinear influence can be neglected as compared with its nonlinear interaction with the mean velocity ${\bm U}$. Since the characteristic scale $1/k_f$ of ${\bm u}$ is assumed to be small, $k_f\rho\gg1$, we introduce local right-handed Cartesian reference $O\rho\varphi z$ which moves and rotates with a Lagrangian particle of the large-scale flow, so $\varphi$-axis is always directed along the streamwise $\varphi$-direction and $\rho$-axis is directed along the radial $\rho$-direction. In the reference system, the dynamics of the turbulent part ${\bm u}$ is described by the Navier-Stokes equation linearized with respect to ${\bm u}$ \cite{kolokolov2020structure},
\begin{equation}\label{eq:u}
    \partial_t {\bm u}
    +
    2[{\boldsymbol\Omega}\times {\bm u}]
    +
    \rho\Sigma\partial_\varphi {\bm u}
    =
    -\nabla p + \nu \Delta {\bm u} + {\bm f},
\end{equation}
where ${\bm f}$ is the driving force and $p$ is the effective pressure.

As the Rossby number for the large-scale flow is assumed to be small, the dynamics of ${\bm u}$ is determined by the Coriolis force in the main approximation, that is by the second term in the left-hand side of (\ref{eq:u}). Thus we deal with an ensemble of inertia waves and expand the velocity field over plane waves with circular polarizations,
\begin{equation}\label{appendix:02}
    {\bm u}(t,{\bm r})
    \ = \
    \mathrm{Re}\sum_{s=\pm1}\itl\left({\mathrm{d}^3k}\right)
    \,a_{{\bf k}s}(t)\,{\bf h}_{\bf k}^s\,
    \exp\big(i({\bf k}\!\cdot\!{\bm r})\big).
\end{equation}
The unit basis vectors ${\bf h}_{\bf k}^s$ are defined in Appendix, see (\ref{hhhhh}), and we use notation for the integration measure in wave-vector space $\itl (d^3k) =  \itl d^3k/\left(2\pi\right)^{3}$. The frequencies of fast harmonic oscillations of expansion coefficients $a_{{\bf k}s}$ are determined by the dispersion law $\omega_{\bf k} = 2\Omega k_z/k$ of inertia waves \cite{greenspan1968theory}.

The influence of the shear leads to the evolution along characteristics
\begin{equation}\label{characteristics}
    {\bf k}^\prime(\tau)
    =
    \left\{ k_\rho + \varsigma \tau k_\varphi, k_\varphi, k_z \right\},
\end{equation}
where $\tau = |\Sigma|t$ is dimensionless time and $\varsigma = \mathop{\mathrm{sign}}\Sigma$. The solution for the wave amplitudes $a_{{\bf k}s}(t)$ at $t=0$ is~\cite{kolokolov2020structure}
\begin{eqnarray}\label{eq:ak}
    a_{{\bf k}s}
    =
    \itl^0_{-\infty}
    \frac{d\tau}{|\Sigma|}
    \sqrt{\frac{k}{k^\prime}}f^s_{k^\prime}(\tau) \exp\big(is\Phi_{{\bf k}}(\tau)-\Gamma(\tau)/2 \big),
\end{eqnarray}
where phase $\Phi_{\bm k}$ describes the oscillation of the waves, so $\partial_\tau\Phi_{\bm k}(\tau) = -2|\Sigma|^{-1}\omega_{{\bf k}^\prime(\tau)}$ in the main approximation at small $\mathrm{Ro}$, $f^s_{k^\prime(\tau)}(\tau)$ is the component of the driving force in the basis ${\bf h}_{\bf k}^s$, and \textquotedblleft viscous exponent\textquotedblright
\begin{eqnarray}\label{Gamma-def}
    \Gamma(\tau)
    =
    \frac{2\nu k_f^2}{|\Sigma|} \itl^0_\tau d\tau_1 k^{\prime 2}(\tau_1).
\end{eqnarray}
In (\ref{Gamma-def}) and below, unless otherwise stated, we use dimensionless wave-vectors ${\bf k}$ that are measured in $k_f$.

The statistics of the turbulent velocity component ${\bm u}$ is assumed to be locally homogeneous in space,  and the inertial waves with the opposite circular polarizations are uncorrelated \cite{kolokolov2020structure}. Hence, the correlation function of the wave amplitude is diagonal in the wave-vector space and in the polarization space,
\begin{eqnarray}\label{aa-corr}
    \langle a_{{\bf q}\sigma}\, a_{{\bf k}s}\rangle
    =
    \frac{\delta({\bf q}+{\bf k})}{(2\pi)^{-3}}\,
    A_{{\bf k} s}\delta^{s\sigma},
    \,\
    A_{{\bf k} s}
    =
    {\mathrm K}_{\bf k}
    -
    \frac{s\mathrm{H}_{\bf k}}{2k_f k}.
\end{eqnarray}
The breaking of the mirror symmetry is determined by the asymmetric part of the correlation function $A_{{\bf k} s}$ with respect to the polarizations. So below we are interested in only the part of $A_{{\bf k} s}$ proportional to $H_{\textbf k}$, the latter is the distribution of mean helicity ${H} = \langle \bm{u}\cdot \bm{\omega} \rangle$ in the wave-vector space, where $\bm{\omega} = \operatorname{curl}\bf{u}$ is the vorticity. The asymmetry is assumed to be created by the driving force that has the following statistics:
\begin{equation}\label{spiral_pump}
    \left\langle f^s_{\bf k}(\tau_1) f^\sigma_{\bf q}(\tau) \right\rangle
    =
    \chi(k)
    \frac{\delta({\bf k} + \bm q)}{(2\pi)^{-3}}
    \frac{\delta(\tau_1 - \tau)}{|\Sigma|^{-1}}
    \left(\epsilon -\frac{s\epsilon_h}{2k_f}\right)
    \delta^{s\sigma}.
\end{equation}
We assume that $\chi(k)$ is isotropic, decays at $k\sim 1$,  tends to zero at $k\to 0$ and is normalized to unity, $\itl(d^3k)\chi(k)=1$. Mean power of the force $\epsilon$ is positive, whereas the mean production rate of helicity
\begin{eqnarray}
    \langle \bm{f}(t) \itl^t_{-\infty}dt^\prime \operatorname{rot} \bm{f}(t^\prime) \rangle = \epsilon_h \itl (d^3k) k \chi(k) \sim \epsilon_h
\end{eqnarray}
may be of any sign. As the random force produces positive power at all wave-vectors, it should be $|\epsilon_h| <2k_f\epsilon$. According to (\ref{eq:ak}), the asymmetric part of the correlation function $A_{\textbf{k}s}$ in $s$ is \begin{eqnarray}\label{amplitude_time}
    \mathrm{H}_{\bf{k}}
    =
    \frac{\epsilon_h}{|\Sigma|}
    \itl^{0}_{-\infty} d\tau\,
    k^\prime\, \chi(k^\prime)\,
    e^{-\Gamma(\tau) } .
\end{eqnarray}

The simplest one-point simultaneous mean which characterizes the symmetry violation is velocity-vorticity correlation function
\begin{equation}\label{uomega}
    H_{ij}
    \equiv
    \langle u_i\omega_j\rangle
    = \frac{1}{2}
    \itl (d^3k) \delta^{\scriptscriptstyle \perp}_{ij} \mathrm{H}_{\bf k},
\end{equation}
where $\delta^{\scriptscriptstyle \perp}_{ij} = \delta_{ij} - k_{i} k_{j}/k^{2}$ is the transverse projector in the Fourier space. The mean (\ref{uomega}) is symmetric, $H_{ij} = H_{ji}$, that is the consequence of the adopted spatial homogeneity.
Similarly to \cite{kolokolov2020structure}, we can obtain expression for the average value of the product of the velocity and vorticity components by changing the order of the integration over $\tau$ and ${\textbf k}$ and then changing integration variable ${\textbf k} \to {\bf q} = {\textbf k}^{\prime}(\tau)$:
\begin{eqnarray}\label{vel_vort_corr}
    H_{ij} =
      \frac{\epsilon_h}{2|\Sigma|}
    \itl^\infty_0 d\tau \itl (d^3q) q \chi(q)
    \delta^{\prime \scriptscriptstyle \perp}_{ij} e^{-\Gamma(\tau)}.
\end{eqnarray}
Dimensionless time $\tau$ here is positive, all characteristics in the integrand have opposite sign with respect to (\ref{characteristics}), notation $\Gamma(\tau)$ implies the same expression as in (\ref{characteristics}) with upper and lower limits $\tau$ and 0 correspondingly. Projector $\delta^{\prime \scriptscriptstyle \perp}_{ij} = \delta_{ij} - q^{\prime}_{i} q^{\prime}_{j}/q^{\prime 2}$, where characteristic ${\textbf q}^{\prime}(\tau)$ is given by formally the same as (\ref{characteristics}) expression. Evaluation of integral (\ref{vel_vort_corr}) involves the integration over time, and the characteristic time is $\tau_* = (|\Sigma|/{2\nu k_f^2})^{1/3}\gg 1$. In terms of previously introduced dimensionless numbers the definition of the new one is $\tau_*^3 \sim \mathrm{Ro}/\mathrm{Ek}$. The viscous exponent is small, $\Gamma\ll1$, at \textquotedblleft ballistic times\textquotedblright\ $\tau\lesssim \tau_\ast$ and is evaluated as $\Gamma(\tau) \approx  \left({\tau}/{\tau_*}\right)^3 {q_\varphi^2}/{3}$ at \textquotedblleft viscous times\textquotedblright\  $\tau\gg \tau_\ast$ and $|q_\varphi|\gtrsim 1/\tau$. At times $\tau\gg \tau_\ast^3$ the viscous exponent $\Gamma\gg1$ for any wave-vector having magnitude $q\sim 1$. First consider mean total helicity ${H} = H_{ii}$. As $\delta^{\prime \scriptscriptstyle \perp}_{ii}=2$, the integral over wave-vectors saturates at $q_{\rho,\varphi,z} \sim 1$. The time integrand is $\sim 1$ up to scale $\tau\sim\tau_*$, after that it diminishes to $\sim(\tau_\ast/\tau)^{3/2}$ due to the viscous exponent. Thus the mean helicity is
\begin{eqnarray}\label{mean_helicity}
  {H} = \itl(d^3q) \mathrm{H}_{{\bf q}}
    \sim
    \frac{\epsilon_h }{|\Sigma|}\tau_*.
\end{eqnarray}
Here and further numerical coefficients are not identified, as they depend also on the specific parameters of the external force correlation function $\chi(q)$.

The main contribution to the helicity (\ref{mean_helicity}) comes from $H_{\varphi\varphi}$ and $H_{zz}$, since corresponding elements of $\delta^{\prime \scriptscriptstyle \perp}_{ij}$ are of the order of unity at large times $\tau\gg 1$. Thus one has estimate $H_{\varphi\varphi} \approx H_{zz} \approx {H}/2$. The first diagonal element of (\ref{vel_vort_corr})  $H_{\rho\rho}$ turns out to be much smaller than the mean helicity ${H}$. Corresponding element of transverse projector is small at $q_\varphi\tau\gg 1$, $\delta^{\prime \scriptscriptstyle \perp}_{\rho\rho}\sim 1/(q_\varphi \tau)^2$. So the main contribution to this element comes from the region $|q_\varphi|\tau \lesssim 1$, where $\delta^{\prime \scriptscriptstyle \perp}_{\rho\rho}\sim 1$. The integral over $q_\varphi$ is accumulated at $|q_\varphi| \sim {1}/{\tau}$. Therefore the time integrand is proportional to ${1}/{\tau}$ up to scale $\tau \sim \tau^3_*$, and the dependence is changed to faster decay {at times $\tau\gg \tau_\ast^3$}. The matrix element is $H_{\rho\rho} \sim (\epsilon_h/|\Sigma|)\ln\tau_*.$

Off-diagonal elements $H_{iz}$ and $H_{zi}$ with $i\neq z$ are equal to zero due to oddness of the integrand over $q_z$. The last off-diagonal element $H_{\rho\varphi}$ contains sign-variable function in the integrand that can be integrated by parts in $\tau$ at fixed $\textbf{q}$:
\begin{eqnarray}
    \itl^\infty_0 d\tau \delta^{\prime\perp}_{\rho\phi}
    e^{-\Gamma(\tau)}
    =
    -\varsigma
    \itl^\infty_0
    \frac{d\tau {q^{\prime 2}\ln q^{\prime 2}} }
        {2\tau_*^{3}}
        e^{-\Gamma(\tau)}
    +
    \varsigma\ln q.
\end{eqnarray}
The main contribution is produced by the first term. The integral is accumulated at $\tau\sim\tau_*$ and is order of $\ln\tau_*$ at $q\sim 1$, that results in $H_{\rho\varphi} \sim -\varsigma\epsilon_h/|\Sigma|\ln\tau_*$.

Finally, we write components of velocity-vorticity correlators in matrix form:
\begin{eqnarray}\label{velocity_vorticity_corr_final}
\hat{H} \sim \frac{\epsilon_h }{|\Sigma|}
\begin{pmatrix}
\ln\tau_* & -\varsigma\ln\tau_* & 0\\
-\varsigma\ln\tau_* & \tau_* & 0\\
0 & 0 & \tau_*
\end{pmatrix}
\end{eqnarray}
Note that only the off-diagonal elements depend on the sign $\varsigma$ of shear rate $\Sigma$. This result is in agreement with the results for velocity-velocity means in three-dimensional geostrophic vortex \cite{kolokolov2020structure}, \cite{ogorodnikov2022structure}, \cite{parfenyev2021velocity}.

\section{The $\alpha$-effect}
\label{sec:alpha}

The evolution of magnetic field is governed by equation \cite{batchelor1950spontaneous}
\begin{eqnarray}\label{twocomponents}
    \partial_t {\bm B}
    = \mathop{\mathrm{curl}}
    \left[\bm v\times {\bm B}\right] + \kappa \Delta {\bm B},
\end{eqnarray}
where $\kappa$ is magnetic diffusion coefficient. Similarly to the velocity field of the liquid, magnetic field is assumed to contain mean component $\overline{\bm B}$ and a fluctuating part $\bm{b}$, $\bm B = \overline{\bm B} + \bm{b}$. The nonzero contributions to the mean electromotive force $\langle[{\bm v}\times {\bm B}]\rangle$ are produced by the interaction between the mean velocity and magnetic fields $\overline{\bm E} = \left[ \bm U \times \overline{\bm B}\right]$ and by the interaction between fluctuating parts of velocity and magnetic fields $\bm{\mathcal{E}} = \langle\left[ \bm u \times \bm b\right]\rangle$. If the scales of the mean fields and the fluctuations are well separated, then ${\bm {\mathcal E}} = \hat \alpha \overline {\bm B}$, where matrix $\hat\alpha$ generally depends on both ${\bm U}$ and $\overline{\bm B}$. We arrive to the equation for the evolution of the mean magnetic field as follows:
\begin{eqnarray}\label{mean_field_component_evolut}
    \partial_t \overline{\bm B}
    = \mathop{\mathrm{curl}}
    \left( \overline {\bm E} + \hat\alpha\overline{\bm B} \right)
    + \kappa\Delta\overline{\bm B},
\end{eqnarray}
To implement the averaging of the electromotive force $\bm{\mathcal{E}}$ over the turbulent pulsations, one needs to find the dynamics of the magnetic field fluctuations ${\bm b}$. We consider the dynamics in the limit of small mean magnetic field $\overline {\bm B}$ so the magnetic field has relatively weak influence on the flow, that is the frequency of Alfven waves $\omega_A \sim k_{f}\left( {\bm B}, {\bf k}\right)/\sqrt{\varrho}$ \cite{alfven1942existence} is much smaller than the frequency of the inertial waves $\omega_{\bf k}$, where $\varrho$ is the mass density of the liquid and ${\textbf k}$ is wave-vector. In other words, Lehnert number is small, $\mathrm{Le} \sim B k/(\sqrt{\varrho} \Omega) \ll 1$. Then the theory developed above for the dynamics of the turbulent pulsations ${\bm u}$ on the background of the strong vortex flow ${\bm U}$ can be extended unchanged for the description of the small-scale magnetic field ${\bm b}$. The dynamics is described by equation (\ref{twocomponents}) linearized in ${\bm b}$ and ${\bm u}$:
\begin{eqnarray}\label{magnetic_start_fluct}
    \partial_t {\bm b}
    +
    (\bm U,\nabla) {\bm b}
    =
    ({\bm b},\nabla)\bm U
    +
    \kappa\Delta{{\bm b}}
    +
    (\overline{\bm B},\nabla)\bm u.
\end{eqnarray}
At arbitrary mean magnetic field, the dynamics of ${\bm u}$ and ${\bm b}$ is coupled. In the limit $\mathrm{Le}\ll1$ considered here, one can find ${\bm b}$ using perturbation theory, so the last term in the right hand side of (\ref{magnetic_start_fluct}) plays role of a non-resonant source for magnetic field ${\bm b}$. Hence, the characteristic scale of ${\bm b}$ coincides with that of ${\bm u}$. Equation (\ref{magnetic_start_fluct}) is written in global reference system. Now we change to the moving local Cartesian reference system $O\rho\varphi z$ to obtain the analogue of (\ref{eq:u}). According to the left-hand side of (\ref{magnetic_start_fluct}), the evolution of ${\bm b}$ occurs along the same characteristics (\ref{characteristics}). Detailed calculations of magnetic field fluctuations and elements of $\hat\alpha$-matrix are made in Appendix. Here we present the key expressions and the final results.

According to the formal solution (\ref{final_mangetic_fluct}) of (\ref{magnetic_start_fluct}), magnetic field ${\bm b}$ is time integral of oscillating inertial wave amplitude with the damping exponent produced by the magnetic diffusion. So, the integral depends on the dimensionless ratio $\tau_{\scriptscriptstyle\Omega} = ({2\Omega}/{\kappa k_f^2})^{1/3}=({\mathrm{Pr}_{\mathrm{m}}}/\mathrm{Ek} )^{1/3}$ of the characteristic velocities. The solution (\ref{final_mangetic_fluct}) provides the connection between $\hat\alpha$-tensor and the velocity-vorticity mean $\hat H$:
\begin{eqnarray}\label{alpha_helicity_relation}
    \alpha_{ip} = \Omega^{-1}
    \itl (d^3k)
    \frac{k^4 \tau_{\scriptscriptstyle\Omega}^{-3}}{k_z^2 + (k/\tau_{\scriptscriptstyle\Omega})^6}
    \left( 2H_{ ip} - {H} \delta_{ip}\right)_{\bf{k}},
\end{eqnarray}
Provided small magnetic Prandtl number, expression is valid if the shear rate is relatively weak ($\Sigma\ll \Omega$ or $\Sigma \ll \kappa k_f^2$). Next we use the solution (\ref{vel_vort_corr}) for $\hat H$ changing the integration variable ${\bf k}$ to ${\bf q}$ and obtain the dependence of $\hat\alpha$ on the statistics of the driving force
\begin{eqnarray}\label{alpha_start_general}
    \alpha_{ip}
    =
    -\frac{\epsilon_h}{\tau_{\scriptscriptstyle\Omega}^3 |\Sigma| \Omega} \itl_0^{\infty} d\tau \itl (d^3q) q \chi(q)\cdot
    \frac{q_i^\prime(\tau) q_p^\prime(\tau) q^{\prime 2}(\tau)}
        {q_z^2 + ( q^{\prime}(\tau)/\tau_{\scriptscriptstyle\Omega})^6 }
    e^{-\Gamma(\tau)},
\end{eqnarray}
that is applicable only if the Rossby number is small, $\mathrm{Ro}\ll1$. The viscous exponent $\Gamma$ in the integrand of (\ref{alpha_start_general}) is determined by dimensionless time $\tau_*$, whereas the denominator in the integrand of (\ref{alpha_start_general}) is determined by dimensionless time $\tau_{\scriptscriptstyle\Omega}$. Therefore expressions for elements of $\alpha$-matrix depend on relation between $\tau_*$ and $\tau_{\scriptscriptstyle\Omega}$, $\tau_*/\tau_{\scriptscriptstyle\Omega} \sim  \left(\mathrm{Ro}/ \mathrm{Pr}_{\mathrm{m}} \right)^{1/3}$. The ratio $\tau_{\scriptscriptstyle\Omega}$ can be both small or big compared to unity, the cases of relatively weak $\tau_{\scriptscriptstyle\Omega}\gg 1$ (that is $\mathrm{Ek}\ll \mathrm{Pr}_\mathrm{m}$) and relatively high Ohmic dissipation $\tau_{\scriptscriptstyle\Omega}\ll 1$ (that is $\mathrm{Ek}\gg \mathrm{Pr}_\mathrm{m}$) should be considered separately. Thus the elements of $\alpha$-matrix sufficiently depend on
relation between $\mathrm{Pr}_{\mathrm{m}}$ and $\mathrm{Ro}$ and between $\mathrm{Pr}_{\mathrm{m}}$ and $\mathrm{Ek}$. Detailed calculations of $\alpha$-matrix's elements are made in Appendix. Here we overview the relation of our calculations to that given in previous works and after that present the final expressions for $\alpha$-matrix at all the considered limiting cases. Neglecting the dependence of the wave-vector on time in (\ref{alpha_start_general}) and thus assuming zero shear, one arrives to the expression obtained in \cite[Eqs.~(3.6,3.8)]{moffatt1970dynamo}. In the limit of weak magnetic diffusion ($\kappa\to 0$), the expression for $\hat{\alpha}$ (\ref{alpha_start_general}) passes into the time integral over different-time velocity-vorticity mean obtained in \cite[Eq.~(3.3)]{moffatt1974mean} and \cite[Eq.~(3.41a)]{radler1980mean} for general-type flows at $\kappa = 0$. For finite magnetic diffusion in the limit of weak rotation, the trace of $\hat{\alpha}$ (\ref{alpha_helicity_relation}) corresponds to the formula obtained in \cite[Eqs.~(3.11,3.13)]{moffatt1970turbulent}, and the tensor itself has the form given in \cite[Eq.~(2.21)]{moffatt1970turbulent}, \cite[Eq.~(10)]{seshasayanan2018growth} and similar to \cite[Eq. above (5.5)]{molchanov1985kinematic} all obtained in the absence of rotation.

First we consider the case of relatively weak Ohmic dissipation (that is $\Omega\gg\kappa k_f^2$ or $\mathrm{Ek}\ll \mathrm{Pr}_\mathrm{m}$). In the limit of fast rotation $\mathrm{Ro}\ll \mathrm{Pr}_{\mathrm{m}}$, the $\hat\alpha$-tensor has the form as follows
\begin{eqnarray}\label{alpha_final_small_Rossby}
    \hat{\alpha} &\sim& -\frac{\epsilon_h }{2\Omega|\Sigma|}
\begin{pmatrix}
 \tau_*^{ 2} & \varsigma\tau_* & 0\\
 \varsigma\tau_* & {a}_{\varphi\varphi} & 0\\
 0 & 0 & {a}_{zz}
 \end{pmatrix},
\end{eqnarray}
where ${a}_{\varphi\varphi} = -\ln(\mathrm{Pr}_{\mathrm{m}}(1+\kappa k_f^2/|\Sigma|))$ and ${a}_{zz} = \mathrm{Ro}/\mathrm{Pr}_{\mathrm{m}}\ll 1$ are both positive. The ratio $|\Sigma|/\kappa k_f^2 = \mathrm{Ro}\mathrm{Pr}_{\mathrm{m}} /\mathrm{Ek}$, and ${a}_{\varphi\varphi}$ depends logarithmically on the flow parameters and thus it is expected to be smaller than the other elements in $\rho\varphi-$plane.  In the absence of shear flow, when $\Sigma\ll\nu k_f^2$, the analogous but axisymmetric anisotropy was established in \cite[Eqs.~(4.1-4.3)]{moffatt1970dynamo}. In the opposite limit $\mathrm{Ro}\gg \mathrm{Pr}_{\mathrm{m}}$, the expression for elements of $\alpha$-matrix is
\begin{eqnarray}\label{alpha_final_bigger_Rossby}
    \hat{\alpha} &\sim& -\frac{\epsilon_h }{2\Omega|\Sigma|}
\begin{pmatrix}
 \tau_{\scriptscriptstyle\Omega}^{2}{a}_{zz} & \varsigma\tau_{\scriptscriptstyle\Omega} & 0\\
 \varsigma\tau_{\scriptscriptstyle\Omega} & {a}_{\varphi\varphi} & 0\\
 0 & 0 & {a}_{zz}
\end{pmatrix},
\end{eqnarray}
where ${a}_{\varphi\varphi} = -\ln(\mathrm{Ro}(1+\kappa k_f^2/|\Sigma|))$ and ${a}_{zz} = \ln\left(\mathrm{Ro}/\mathrm{Pr}_{\mathrm{m}}\right)$ are again both positive and larger than unity, but expected to be smaller than $\tau_{\scriptscriptstyle\Omega}$. As calculations were made with logarithmic accuracy, one solution (\ref{alpha_final_small_Rossby}) turns to the other (\ref{alpha_final_bigger_Rossby}) at $\mathrm{Ro}\sim \mathrm{Pr}_{\mathrm{m}}$. Up to the logarithmic corrections, the expressions (\ref{alpha_final_small_Rossby}) and (\ref{alpha_final_bigger_Rossby}) projected on the $\rho\varphi$-plane can be joined if one takes $\mathop{\mathrm{min}}(\tau_\ast,\tau_{\scriptscriptstyle\Omega})$. The projection weakly depends on the magnetic diffusion in the limit (\ref{alpha_final_small_Rossby}) and starts to decrease with the increasing diffusion in the limit (\ref{alpha_final_bigger_Rossby}). It can be notices also that the magnitude of $\hat \alpha$ decreases with the increasing rotation.

In the case of the relatively high Ohmic dissipation (that is $\Omega\ll\kappa k_f^2$ or $\mathrm{Ek}\gg \mathrm{Pr}_\mathrm{m}$), the characteristic time needed for magnetic field ${\bm b}$ is formed by the velocity fluctuations ${\bm u}$ is now magnetic diffusive time, so $\alpha$ has following form:
\begin{eqnarray}\label{alpha_final_prandtl_smaller_ekman}
    \hat{\alpha} \sim -\frac{\epsilon_h}{\kappa k_f^2|\Sigma|}
    \begin{pmatrix}
        \ln\tau_* & \varsigma & 0\\
        \varsigma & 1 & 0\\
        0 & 0 & \ln\tau_*
    \end{pmatrix}.
\end{eqnarray}
The magnitude of $\hat \alpha$ (\ref{alpha_final_prandtl_smaller_ekman}) decreases with the increasing magnetic diffusion coefficient and does not depend on the rotation frequency.

As calculations were made with logarithmic accuracy, one solution (\ref{alpha_final_prandtl_smaller_ekman}) turns to the other (\ref{alpha_final_bigger_Rossby}) at $\mathrm{Ek}\sim \mathrm{Pr}_{\mathrm{m}}$ (that is at $\tau_{\scriptscriptstyle\Omega}\sim 1$). Due to the symmetry reasoning, off-diagonal elements always depend on the sign of the shear rate $\Sigma$ while diagonal elements do not depend on the sign. It follows from the presented calculations, that $\alpha_{\rho\rho}$, $\alpha_{\rho\varphi}$ and $\alpha_{zz}$ are accumulated at $q_z\gg \mathrm{Ro}$. However, $\alpha_{\varphi\varphi}$ is determined by an integral which logarithmically diverges in $q_z$. The cases (\ref{alpha_final_small_Rossby},\ref{alpha_final_bigger_Rossby}) are formed by  $\mathop{\mathrm{max}}(\mathrm{Ro},\kappa k_f^2/\Omega)\lesssim q_z\lesssim \mathop{\mathrm{min}}(\mathrm{Ro}/{\mathrm{Pr}}_\mathrm{m},1)$. The lower restriction $q_z\gtrsim \mathrm{Ro}$ is due to the value of $\alpha$-effect turns to zero in the limit of pure two-dimensional case $q_z=0$ at small Lehnert number $\operatorname{Le}\ll 1$, that can be straightforwardly checked using the same technique applied in \cite{parfenyev2021influence}. Our theory used here is applicable for $q_z \sim \mathrm{Ro}$ as well. The logarithmic type of the divergency is the consequence of assumed isotropy of the force correlation function $\chi({\bf q})$. However it is not isotropic in real systems and accumulated at smaller $q_z$ \cite{davidson2014dynamics}. In this case, the value of $\varphi\varphi-$element of $\alpha$-matrix is determined by the vicinity of the smallest $q_z\sim \mathrm{Ro}$. There is a relation between the value of $q_z$ and the characteristic value of time $\tau$ for the integrals, $\tau \sim \tau_{\scriptscriptstyle\Omega} q_z^{1/3}$, so the interval for $q_z$ corresponds, in particular, to $\mathop{\mathrm{max}}(\tau_{\scriptscriptstyle\Sigma},1)\lesssim \tau\lesssim \mathop{\mathrm{min}}(\tau_\ast, \tau_{\scriptscriptstyle\Omega})$ where we use notation $\tau_{\scriptscriptstyle\Sigma} = (|\Sigma|/\kappa k_f^2)^{1/3}$.

\section{Mean magnetic field}
\label{sec:mean-magnetic-field}

Here we consider the initial stage of growth of mean magnetic field and establish the threshold condition for the kinematic dynamo. As the vortex is assumed to be strong, we assume that the mean magnetic field is axially symmetric and depends on radial and vertical coordinates only, $\overline{\bm B} = \overline{\bm B}(\rho,z)$. Then the curl of the electromotive force $\overline{\bm E}$ in (\ref{mean_field_component_evolut}) is  $\mathop{\mathrm{curl}}\overline{\bm E} = \Sigma \overline B^\rho {\bf e}_\varphi$. Since we study evolution of the large-scale magnetic field, we should consider dynamics of magnetic field at wavenumbers $K \sim {K}_b \ll k_f$. Here and further we assume that the wave-vectors have the dimension of the inverse length. Consider the case when the characteristic spatial scale of the mean field is small compared to the size of the vortex $R_{u}$, ${K}_b R_{u}\gg 1$, then any spatial derivative of magnetic field can be written in local Cartesian system $O\rho \varphi z$. According to the results presented in Section~\ref{sec:alpha}, the inequality $|k_f \alpha_{ij}| \ll |\Sigma|$ is always valid for all matrix elements. Keeping only the most significant terms in the limit $R_u^{-1}\ll {K}_b \ll k_f$, we arrive to the following equation for the mean magnetic field in Fourier space
\begin{eqnarray}\label{large_magnetic_Fourier_simple}
\partial_t
\begin{pmatrix}
  \overline{B}_{\scriptscriptstyle {\bf K}}^\rho\\
  \overline{B}_{\scriptscriptstyle {\bf K}}^\varphi\\
 \overline{B}_{\scriptscriptstyle {\bf K}}^z
\end{pmatrix}
 \approx
 \begin{pmatrix}
  - \kappa {K}^2 & -i{K}_z\alpha_{\varphi\varphi}  & 0\\
  \Sigma & - \kappa {K}^2 & 0\\
  0 & 0 & - \kappa {K}^2\\
 \end{pmatrix}
 \begin{pmatrix}
 \overline{B}_{\scriptscriptstyle {\bf K}}^\rho\\
 \overline{B}_{\scriptscriptstyle {\bf K}}^\varphi\\
 \overline{B}_{\scriptscriptstyle {\bf K}}^z
\end{pmatrix}
\end{eqnarray}
The linear system (\ref{large_magnetic_Fourier_simple}) has one eigenvalue $\lambda$ which may have positive real part:
\begin{eqnarray}\label{eigenvalues_final}
    \lambda
    \approx
    -\kappa {K}^2
    +
    \sqrt{\frac{|\alpha_{\varphi\varphi}{K}_z \Sigma|}{2}}\left( 1 + i\cdot \mathrm{sign}\left(\epsilon_h {K}_z \Sigma\right)\right).
\end{eqnarray}
It has positive real part only if $\sqrt{|\alpha_{\varphi\varphi} {K}_z \Sigma|/2}>\kappa {K}^2$. According to (\ref{eigenvalues_final}) it is the $\varphi\varphi$-element that determines the criterion for the growth of a large-scale magnetic field, and its geometrical role is transformation of a poloidal magnetic field into toroidal one \cite{brandenburg1997dependence}. This key role of the element was justified earlier for a general type flows that are close to axially symmetric \cite[after (6)]{braginsky1990some}. Similar criterion was obtained for the magnetic field $\overline{\bm B}$ that depends only on the coordinate along the axis of rotation $z$ in \cite{brandenburg2002local}, here it is generalized for an arbitrary radial dependence. At finite ${K}_\rho$ in our case, the threshold can be overcome if $|{K}_\rho|<\sqrt{3} K_\star$ with $K_\star=2^{-5/3}(|\alpha_{\varphi\varphi}\Sigma|/\kappa^2)^{1/3}$, then the magnetic field harmonic with the greatest growth rate  $\lambda=\kappa(3K_\star^2-{K}_\rho^2)$ has ${K}_z=K_\star$. It can be also checked that the rate $\lambda$ is always slow in a sense $|\lambda|\ll\Omega$.

Next we establish a criterion for the magnetic field growth in terms of non-dimensional parameters and determine the configuration of the large-scale magnetic field. We assume that the shear rate is maintained by the power supplying by the small-scale turbulent  pulsations, $\epsilon\sim \nu\Sigma^2$ \cite{kolokolov2020structure}, and the degree of the mirror symmetry breaking is of the order of unity, $|\epsilon_h| \sim \epsilon k_f$. In the limit of fast rotation $\Omega\gg \kappa k_f^2$  (that is $\mathrm{Ek}\ll\mathrm{Pr}_\mathrm{m}$, when (\ref{alpha_final_small_Rossby},\ref{alpha_final_bigger_Rossby}) are applicable), the criterion has following form:
\begin{eqnarray}\label{restriction}
    \frac{\mathrm{Ek}}{\mathrm{Pr}_{\mathrm{m}}} \lesssim
    \mathrm{Ro}^2 \mathrm{Pr}_{\mathrm{m}} \left(\frac{k_f}{{K}_b}\right)^3\frac{\epsilon}{\nu\Sigma^2},
\end{eqnarray}
where we did not take into account logarithm in $\alpha_{\xi\xi}$ and adopted ${K}_z\sim {K}_b$. In the limit of high Ohmic dissipation $\Omega\ll \kappa k_f^2$ (that is $\mathrm{Ek}\gg\mathrm{Pr}_\mathrm{m}$, when (\ref{alpha_final_prandtl_smaller_ekman}) is applicable) the criterion is
\begin{eqnarray}\label{restriction_second}
    \left(\frac{\mathrm{Ek}}{\mathrm{Pr}_{\mathrm{m}}}\right)^2 \lesssim
    \mathrm{Ro}^2 \mathrm{Pr}_{\mathrm{m}} \frac{\epsilon}{\nu\Sigma^2}\left(\frac{k_f}{{K}_b}\right)^{3}.
\end{eqnarray}

Let's now find eigenvector $\overline {\bm B}^\lambda_{\scriptscriptstyle{\bf K}}$ that corresponds to eigenvalue $\lambda$ taking into account that the magnetic field is solenoidal, ${K}_\rho \overline B^{\lambda,\rho}_{\scriptscriptstyle{\bf K}} + {K}_z \overline B^{\lambda, z}_{\scriptscriptstyle{\bf K}} =0$. We assume that the dynamo threshold is significantly overcome, so the first term $-\kappa {K}^2$ in (\ref{eigenvalues_final}) is relatively small. Then the complex amplitude of the magnetic field is
\begin{eqnarray}\label{configuration_simple}
    \overline{\bm B}^\lambda_{\scriptscriptstyle{\bf K}}
    \propto
\begin{pmatrix}
  1 \\
  -|\Sigma|/\lambda
  \\
  -{K}_\rho/{K}_z
\end{pmatrix} e^{\lambda t},
    \quad
    \frac{|\Sigma|}{|\lambda|}
    \sim
    \frac{\sqrt{k_f/{K}_b} \sqrt{\epsilon/(\nu \Sigma^2)}}{\sqrt{\mathop{\mathrm{min}}({\mathrm{Ek}},{\mathrm{Pr}_\mathrm{m}})}}
    \gg 1.
\end{eqnarray}
The azimuth component of the field is relatively large as $|\Sigma/\lambda|$, so the magnetic field is mainly directed along azimuth $\varphi$-axis. Also note that the $\rho$-component is ahead of $\varphi$-component in phase by {$3\pi/4$} if $\epsilon_h {K}_z\Sigma>0$, otherwise the phase difference has the inverse sign.

\section{Discussion}
\label{sec:discussion}

The presented calculations in the previous Section assume that the shear flow is unbounded and uniform. The finite scale $R_u$ of the large-scale flow leads to the inertia waves may pass the whole scale before they are absorbed by the flow. The group velocity of the waves is $\sim \Omega/k_f$, so the dimensionless passage time can be evaluated as $\tau_t=
|\Sigma| R_u/(\Omega/k_f)$. At times $\tau \gtrsim \tau_{t}$ developed theory is not applicable and $\tau_{t}$ becomes the maximum possible upper limit in all time integrals \cite{kolokolov2020structure, ogorodnikov2022structure}, so calculations of tensors $\hat H$ and $\hat\alpha$ should be revisited. Here we estimate $\alpha_{\varphi\varphi}$ because only this element leads to the magnetic field generation. In the limit $\mathrm{Ek} \gg \mathrm{Pr}_\mathrm{m}$ integral (\ref{detailed_tau_beta_xixi}) is accumulated at $q\sim 1$ and $\tau \sim 1$ thus in this limit element $\alpha_{\varphi\varphi}$ does not depend on $\tau_{t}$ and remains the same as in the limit $\tau_{t} \to \infty$. The limit $\mathrm{Ek} \ll \mathrm{Pr}_\mathrm{m}$ is less trivial. First we note that the expression (\ref{alpha_final_small_Rossby},\ref{alpha_final_bigger_Rossby}) for $\alpha_{\varphi\varphi}$ depends on $\mathop{\mathrm{min}}\left\{ \tau_{\scriptscriptstyle\Omega}, \tau_* \right\}$ in the limit $\tau_{t} \to \infty$. In the case $\tau_{\scriptscriptstyle\Sigma} \ll \tau_{t} \ll \mathop{\mathrm{min}}\left\{ \tau_{\scriptscriptstyle\Omega}, \tau_* \right\}$ this element has also logarithmic behaviour and can be expressed via dimensionless quantity ${a}_{\varphi\varphi} = -\ln\big( \tau_{t}^{-3}\big(1+|\Sigma|/(\kappa k_f^2) \big) \big)$.
In the case $1\ll \tau_{t} \ll \tau_{\scriptscriptstyle\Sigma}$ the expression depends on upper time limit $\tau_{t}$ according to the power law and can be expressed via dimensionless quantity ${a}_{\varphi\varphi} \sim \left(\tau_{t}/\tau_{\scriptscriptstyle\Sigma} \right)^{3}$.

Both the large-scale shear flow on the background of fast rotation and the small-scale turbulence are assumed in our model. The considered dynamo is allowed if one of conditions (\ref{restriction},\ref{restriction_second}) is met and its mechanism is called $\alpha\Omega-$dynamo generation \cite{tobias2021turbulent, sokoloff2014dynamos}. It is interesting to check the applicability of the obtained results to known experiments. Dynamo is a threshold phenomenon, so the main efforts in the experiments are aimed at exceeding the thresholds for magnetic Reynolds number $\mathrm{Re}_\mathrm{m}^*$ that is usually $\mathrm{Re}_\mathrm{m}^* \sim 10 \div 100$ \cite{sokoloff2014dynamos}. The experiments were carried out predominantly with liquid sodium that has $\kappa \sim 0.1 m^2/s$, $\nu \sim 10^{-6} m^2/s$, so magnetic Prandtl number is small, $\mathrm{Pr}_\mathrm{m} \sim 10^{-5}$. The first dynamo-experiment was made in Riga \cite{gailitis2000detection} where so-called Ponomarenko dynamo \cite{ponomarenko1973theory} was realised.
In this case only large-scale helical flow is responsible for magnetic field generation. A non-stationary helical flow in toroidal channel was achieved in Perm experiments \cite{frick2002non, frick2012turbulent, sukhanovskii2023unsteady}, so the generated magnetic field was non-stationary as well. Another dynamo experiment was done in Karlsruhe \cite{stieglitz2001experimental} where the double-scale dynamo was obtained. The $\alpha\Omega-$dynamo was realised in the experiment in Cadarache \cite{monchaux2007generation}. However in this experiment dynamo was obtained for von Karman swirling flow that differs from our case of geostrophic flow.

So, the flow obtained in the Perm experiments \cite{frick2012turbulent} comes the closest to our problem formulation. The flow with large enough Reynolds number ($\mathrm{Re}\sim 3\cdot 10^6$) was accelerated in toroidal channel and then was braked up abruptly. We can extract following data from experiments \cite{frick2012turbulent}. The global rotation frequency is $\Omega \sim 300 s^{-1}$, the scale $k_f$ can be estimated $k_f \sim N/r_0$, where $r_0$ is the radius of the cross-section of toroidal channel, $N = 10$ is the amount of blades in the divertor. For the MHD apparatus \cite{frick2012turbulent} we have $r_0 \sim 0.08 m$, that gives Ekman number $\mathrm{Ek} \sim 10^{-5}$, so we have $\mathrm{Ek}\sim \mathrm{Pr}_\mathrm{m}$ and should check the magnetic field instability criterion in form (\ref{restriction}). If the mean magnetic field achieves the largest available scale, then the ratio $k_f/{K}_b\sim N = 10$. We can adopt $\mathrm{Ro}\lesssim 1$, relative helicity is of the order of unity, $\epsilon_h \sim \epsilon k_f$, and the differential rotation in the vortex is supported by upscale energy transfer,  $\epsilon \sim \nu \Sigma^2$ \cite{kolokolov2020structure}. Thus the right hand side of (\ref{restriction}) is order of $10^{-2}$. It is actually smaller than the left hand side $\sim 1$, so the criterion is not met. Note that there are no two different scales of large-scale flow and turbulent pulsations in observed dynamo \cite{frick2012turbulent}, so this dynamo can be called quasi-laminar in terms of \cite{sokoloff2014dynamos}. This is the reason why considered criterion is not applicable for the dynamo in the experiments.

To summarise, we have quantitatively characterized $\alpha$-effect in three-dimensional coherent geostrophic helical vortex of conducting liquid with external magnetic field. Under assumption that there is an external volume forcing acting on the flow and possessing non-zero helicity, we have calculated elements of one-point velocity-vorticity mean $\langle u_i \omega_k\rangle$ for the fluctuating part of the flow and $\alpha$-matrix that describes magnetic dynamo $\alpha$-effect. The results were obtained in considered region $\mathrm{Ro}\lesssim 1, \,\  \mathrm{Pr}_{\mathrm{m}} \lesssim 1$, the choice of the range corresponds to three-dimensional vortex flow and liquid metal. The calculations were based on the theory developed in \cite{kolokolov2020structure} and \cite{ogorodnikov2022structure}. The elements of $\alpha$-matrix strongly depends on relation between Rossby $\mathrm{Ro}$ and magnetic Prandtl $\mathrm{Pr}_{\mathrm{m}}$ numbers. After calculating the pseudotensor $\hat{\alpha}$, the criterion (\ref{restriction}) for growth of large-scale magnetic field was established.

\section{Acknowledgments.} This work was supported by the {Russian Science Foundation, Grant No. 20-12-00383.}

\section{Appendix: Calculation of the $\alpha-$effect}\label{appendix_first}

In the Appendix, we present the calculations of $\hat\alpha$-matrix, which determines the electromotive force $\bm{\mathcal{E}} = \langle\left[ \bm u \times \bm b\right]\rangle = \hat\alpha \overline {\bm B}$ (\ref{mean_field_component_evolut}). To find the mean, one should express the magnetic field fluctuations ${\bm b}$ via turbulent pulsations ${\bm u}$ using equation (\ref{magnetic_start_fluct}) \cite{steenbeck1966berechnung}, \cite{braginskii1964theory}, \cite{braginsky1990some}. We change to moving local reference frame $O\rho\varphi z$, approximate the large-scale flow by shear flow, use the dimensionless time $\tau = |\Sigma|t$ and pass into momentum space:
\begin{eqnarray}\label{magnetic_kspace}
    |\Sigma|\frac{d {\bm b}_{{\bf k}^\prime}}{d \tau}
    -
    \Sigma b_{{\bf k}^\prime}^\rho \bm e_\varphi
    +
    \kappa k_f^2 k^{\prime 2}{{\bm b}}_{{\bf k}^\prime}
    =
    i k_f\overline{B}^p k_p^\prime \bm u_{{\bf k}^\prime}
\end{eqnarray}
Equation (\ref{magnetic_kspace}) sets the movement along characteristics (\ref{characteristics}). Next we move to the basis of orthogonal polarisation ${\bf h}^s_{\bf k}$:
\begin{eqnarray}\label{hhhhh}
    {{\bm b}}_{{\bf k}^\prime} = \sum_s c_{{\bf k}^\prime s} {\bf h}^s_{{\bf k}^\prime}, \qquad
    {\bf h}_{\bf k}^s
    =
    \frac{[{\bf k}\times[{\bf k}\times {\bf e}_z]]+ isk[{\bf k}\times {\bf e}_z]}{\sqrt{2}k^2\sin\theta_{\bf k}},
    \qquad s = \pm 1.
\end{eqnarray}
The evolution of the expansion coefficients $c_{{\bf k}^\prime s}$ is governed by
\begin{eqnarray}\label{eq:100}
    \left\{
        |\Sigma|\left({\bf h}^{-s}_{{\bf k}^\prime}, \frac{d {\bf h}^{s}_{{\bf k}^\prime}}{d\tau}  \right)
        -
        \Sigma  h^{s,\rho}_{{\bf k}^\prime}  h^{-s,\varphi}_{{\bf k}^\prime}
        +
        \kappa k_f^2 k^{\prime 2}
        \right\}
    c_{{\bf k}^\prime s}
    +
    |\Sigma|\frac{d c_{{\bf k}^\prime s}}{d \tau} = i k_f\left( {\bf h}^{-s}_{{\bf k}^\prime}, \bm u_{\bf{k}^\prime}\right) k^\prime_j \overline{B}^j
\end{eqnarray}
Without loss of generality, we adopt $\tau=0$, then the solution for  (\ref{eq:100}) is
\begin{eqnarray}\label{final_mangetic_fluct}
    b^i_{\bf{k}}
    =
    ik_f|\Sigma|^{-1}
    \overline{B}^j
    \sum_s h^{s,i}_{{\bf k}}
    \itl^0_{-\infty} dT k^\prime_j(T) a_{\bf{k}^\prime s}(T)
    \sqrt{\frac{k}{k^\prime(T)}}
    \exp\left(-\frac{\Gamma(T)}{2 \mathrm{Pr}_{\mathrm{m}}}  + is \phi_{{\bf k}}(T) \right)
\end{eqnarray}
where the velocity expansion coefficients $a_{{\bf k},s}$ are defined in (\ref{appendix:02}). As it will be shown below, the electromotive force $\bm{\mathcal{E}}$ is determined by $T\sim \mathrm{Ro}$, so we have neglected phase multiplier in (\ref{final_mangetic_fluct}) that changes at times $T\sim 1$. According to (\ref{final_mangetic_fluct}), the calculation of electromotive force $\bm{\mathcal{E}}$ needs in different-time correlation function of the expansion coefficients
\begin{eqnarray}\label{time-dependent-correl}
    \big\langle  a_{\bf{k}^\prime s}(0)a_{\bm{p}^\prime(T)\sigma}(T)\big\rangle
    =
    (2\pi)^3 \delta({\bf k}+{\bf p}) \delta^{s\sigma}
    A_{{\bf k}^\prime s}
    e^{is\Phi_{{\bf k}}(T)},
\end{eqnarray}
Here wave spectrum $A_{\bf{k}^\prime s}$ and phase $\Phi_{{\bf k}}(T)$ are defined in (\ref{aa-corr}) and after (\ref{eq:ak}) respectively, and we have neglected both the viscosity and the movement along the characteristics (\ref{characteristics}) as their influence is weak at $T\ll1$. Using relation (\ref{time-dependent-correl}), we can express the electromotive force via velocity flow (\ref{appendix:02}) and magnetic field (\ref{final_mangetic_fluct}) fluctuations:
\begin{eqnarray}\label{electro_force}
    \mathcal{E}_i
    =
    -\frac{\overline{B}^j }{\Sigma^2} \itl(d^3k)
    \frac{k_i k_j}{k^2}
    {\mathrm H}_{\bf{k}}
    \mathrm{Re}
    \itl^0_{-\infty} dT
    \exp\left(-\frac{\Gamma(T)}{2 \mathrm{Pr}_{\mathrm{m}}} +i\Phi_{{\bf k}}(T)\right).
\end{eqnarray}
Now we substitute ${\mathrm H}_{\bf k}$ (\ref{amplitude_time}) in  (\ref{electro_force}) and change wave-vector ${\bf k}$ to ${\bf q}$ inside the integral, see before (\ref{vel_vort_corr}), after that we arrive to (\ref{alpha_start_general}). In the sake of brevity, further it is convenient to introduce the dimensionless quantity ${a}_{ij} =  -|\Sigma|2\Omega\alpha_{ij}/\epsilon_h$:
\begin{eqnarray}\label{beta_start_general_0}
    {a}_{ij}
    =
    \frac{\kappa k_f^2}{2\Omega} \itl_0^{\infty} d\tau \itl (d^3q) q \chi(q)\cdot
    \frac{q_i^\prime(\tau) q_j^\prime(\tau) q^{\prime 2}(\tau)}
        {q_z^2 + ( q^{\prime}(\tau)/\tau_{\scriptscriptstyle\Omega})^6 }
    e^{-\Gamma(\tau)}.
\end{eqnarray}
One can immediately notice that ${a}_{z\varphi}={a}_{z\rho}=0$ since the integrand is odd in $q_z$. Next, all diagonal elements ${a}_{ii}$ are positive because numerator in the integrand in (\ref{alpha_start_general}) contains $q_i^{\prime 2}(\tau)$ that is always positive. The only nonzero off-diagonal element, ${a}_{\varphi\rho}$, contains the product $q_\varphi q^\prime_\rho(\tau)$ in the integrand in (\ref{alpha_start_general}), so the expression is not sign-determined. Therefore some approaches for the diagonal elements are not applicable for ${a}_{\varphi\rho}$.

\subsection{Case $\mathrm{Pr}_\mathrm{m}\gg\mathrm{Ek}$}

The case corresponds to $\tau_{\scriptscriptstyle\Omega}\gg 1$.
Let's start from diagonal elements. For all these elements integral (\ref{alpha_start_general}) over $q_\rho$ is accumulated at  $q_\rho \sim 1$ due to the structure of $\chi(q)$. Further we need to take integrals over $q_z$, $q_\varphi$ and $\tau$ successively. Note that spectrum behaviour in two-dimensional sector ($q_z/q \lesssim \mathrm{Ro}$) differs from the three-dimensional one ($q_z/q \gtrsim \mathrm{Ro}$) due to smaller contribution from inertial waves that oscillate with frequency $\omega_{{\bf q}}$ \cite{parfenyev2021influence}. Therefore the developed three-dimensional theory is not applicable in two-dimensional region. Thus we should check performing calculations at the three-dimensional sector that the formal contribution to (\ref{alpha_start_general}) from two-dimensional sector remains negligible for our calculations to be reasonable. Furthermore, ${a}_{zz}$ is determined at moderate $q_z\sim 1$ for all values of parameters but ${a}_{ij}$ in $\rho\varphi$-plane can be determined by small $q_z$ at some values of Rossby $\mathrm{Ro}$ and Prandtl magnetic $\mathrm{Pr}_{\mathrm{m}}$ numbers. Therefore we consider ${a}_{zz}$ and ${a}_{ii}$ in $\rho\varphi-$ plane separately.

\paragraph{Calculation of zz-element.} The element is accumulated at times $\tau\gg 1$ so after substitution $q^\prime_\rho \to \varsigma k_\varphi \tau$ it will have following form:
\begin{eqnarray}\label{betazzstart}
    {a}_{zz} \sim \tau_{\scriptscriptstyle\Omega}^{-3}
    \itl^{\infty}_1 d\tau \tau^2 \itl (dq_\varphi) q_\varphi^2
    \itl^\infty_{\mathrm{Ro}}  \frac{(dq_z) q_z^2 }{q_z^2 + \left(q_\varphi {\tau}/{\tau_{\scriptscriptstyle\Omega}}\right)^6} \itl (dq_\rho) q\chi(q) e^{-\Gamma(\tau)}
\end{eqnarray}
Here and further numerical coefficients are not identified, as they depend also on the specific parameters of the external force
correlation function $\chi(q)$. The integral over $q_{z,\rho}$ is determined by $q_{z,\rho} \sim 1$. The region where integral over other variables $q_\varphi$ and $\tau$ is saturated is determined by denominator, $|q_\varphi| \lesssim \tau_{\scriptscriptstyle\Omega}/\tau$, viscous exponent $\Gamma$, $|q_\varphi| \lesssim (\tau_*/\tau)^{3/2} $, and external force
correlation function $\chi(q)$, $|q_\varphi| \lesssim 1$. Therefore the integration over the variables $q_\varphi$ and $\tau$ significantly depends on the relation between $\tau_*$ and $\tau_{\scriptscriptstyle\Omega}$, i.e. between $\mathrm{Ro}$ and ${\mathrm{Pr}}_m$.

In the limit $\mathrm{Ro}\ll \mathrm{Pr}_{\mathrm{m}}$, i.e. $\tau_*\ll\tau_{\scriptscriptstyle\Omega}$, integral (\ref{betazzstart}) is accumulated at $\tau\sim\tau_*$ and $q_\varphi \sim 1$.
After the integration over ${\bm q}$, the integrand has the form $\sim \tau^2/\tau_{\scriptscriptstyle\Omega}^3$ at $\tau\lesssim \tau_*$ and rapidly decreases with time as $\sim \sqrt{{\tau_*^{9}}/(\tau^{5}\tau_{\scriptscriptstyle\Omega}^6)}$ at $\tau\gtrsim \tau_*$. Thus zz-element in the limit $\mathrm{Ro}\ll \mathrm{Pr}_{\mathrm{m}}$ is
\begin{eqnarray}\label{zz01}
    {a}_{zz} \sim \frac{\mathrm{Ro}}{\mathrm{Pr}_{\mathrm{m}}}.
\end{eqnarray}
In the limit $\mathrm{Ro}\gg \mathrm{Pr}_{\mathrm{m}}$ integral is accumulated at $q_\varphi \sim {\tau_{\scriptscriptstyle\Omega}}/{\tau}$ and $\tau_{\scriptscriptstyle\Omega} \lesssim \tau \lesssim \tau_*^3/\tau_{\scriptscriptstyle\Omega}^2$. The integrand has the form $\sim \tau^2/\tau_{\scriptscriptstyle\Omega}^3$ at $\tau\lesssim\tau_{\scriptscriptstyle\Omega}$, slowly decreases as $\sim {1}/{\tau}$ at $\tau_{\scriptscriptstyle\Omega} \lesssim \tau \lesssim \tau^3_*/\tau_{\scriptscriptstyle\Omega}^2$ and rapidly decreases as $\sim \sqrt{{\tau_*^{9}}/(\tau^{5}\tau_{\scriptscriptstyle\Omega}^6)}  $ at $\tau \gtrsim \tau_*^3/\tau_{\scriptscriptstyle\Omega}^2$ after the integration over all wave-vector's components. Therefore the main contribution to the integral has logarithmic behaviour:
\begin{eqnarray}\label{zz02}
    {a}_{zz} \sim \ln\left(\frac{\mathrm{Ro}}{\mathrm{Pr}_{\mathrm{m}}}\right).
\end{eqnarray}

\paragraph{Calculation of $\rho\rho$- and $\varphi\varphi$-elements.}
Again, times $\tau\gg 1$ are relevant, so we substitute $q^\prime_\rho \to \varsigma k_\varphi \tau$ and have:
\begin{eqnarray}\label{xi_eta_start}
    {a}_{ij} \sim  \tau_{\scriptscriptstyle\Omega}^{-3}
    \itl^\infty_{\sim 1} d\tau \tau^{2+n_{ij}}
    \itl_{\sim 1/\tau}^{\infty}(d q_\varphi) q_\varphi^4
    \itl^\infty_{\sim\mathrm{Ro}} \frac{(dq_z) }{q_z^2+\left(q_\varphi {\tau}/{\tau_{\scriptscriptstyle\Omega}}\right)^6} \itl (dq_\rho) q\chi(q) e^{-\Gamma(\tau)},
\end{eqnarray}
where $n_{\rho\rho}=2$, $n_{\varphi\varphi}=0$. The restrictions imposed on $k_\varphi$-region where the integral is saturated are the same as in the case of $zz$-element.  We remind that it was defined $\tau_{\scriptscriptstyle\Sigma}\ll \tau_\ast, \tau_{\scriptscriptstyle\Omega}$ at the very end of Section~\ref{sec:alpha}. Thus integral over $q_z$ in (\ref{xi_eta_start}) is accumulated at $q_z \sim \left( q_\varphi \tau/\tau_{\scriptscriptstyle\Omega}\right)^3 \ll \mathrm{Ro}$ at times $\tau\ll \tau_{\scriptscriptstyle\Sigma}$, that corresponds to two-dimensional sector where developed in (\cite{ogorodnikov2022structure}) three-dimensional theory is not applicable. It can be easily checked that the formal contribution to (\ref{xi_eta_start}) from times $\tau\lesssim\tau_{\scriptscriptstyle\Sigma}$ remains negligible so our calculations are reasonable.

\textit{Let's first calculate ${a}_{\rho\rho}$.} In the limit $\mathrm{Ro}\ll \mathrm{Pr}_{\mathrm{m}} $, integral (\ref{xi_eta_start}) is determined by $q_\varphi \sim 1,$ $q_z \sim \left(\tau_*/ \tau_{\scriptscriptstyle \Omega}\right)^3$ and $\tau\sim\tau_*$ correspondingly. Note that at typical scale of $q_z \sim \mathrm{Ro}/\mathrm{Pr}_\mathrm{m}\gg \mathrm{Ro}$ inertial waves are still relatively rapid. The integrand depends on time as $\sim \tau^{2}$ at $\mathop{\mathrm{max}}\left\{ 1, \tau_{\scriptscriptstyle\Sigma} \right\}\lesssim\tau\lesssim \tau_*$ and as $\sim { \tau^3_*}/{\tau^{2}}$ at $\tau\gtrsim \tau_*$ after the integration over all components of wave-vector. So ${a}_{\rho\rho} \sim \tau_*^2$.

In the opposite limit, $\mathrm{Ro}\gg \mathrm{Pr}_{\mathrm{m}}$, the main contribution to the integral is given by $q_z\sim 1,$ $q_\varphi \sim \tau_{\scriptscriptstyle\Omega}/\tau$ and $\tau_{\scriptscriptstyle\Omega} \lesssim \tau \lesssim \tau_*^3/\tau_{\scriptscriptstyle\Omega}^2$. The viscous exponent has the form $\Gamma \sim q_\varphi^2 (\tau/\tau_*)^3  \sim \tau_{\scriptscriptstyle\Omega}^2 \tau/\tau_*^3 $ at these wave-vectors and times. The integrand depends on time as $\sim \tau$ at $\mathop{\mathrm{max}}\left\{1, \tau_{\scriptscriptstyle\Sigma} \right\} \lesssim \tau\lesssim \tau_{\scriptscriptstyle\Omega}$, as $\sim \tau_{\scriptscriptstyle\Omega}^2/\tau$ at $\tau_{\scriptscriptstyle\Omega}\lesssim \tau\lesssim \tau_*^3/\tau_{\scriptscriptstyle \Omega}^2$ and as $\sim \tau_*^3/\tau^{2}$ at larger times after the integration over all wave-vector's components. The main contribution is accumulated at range $\tau_{\scriptscriptstyle\Omega}\lesssim \tau\lesssim \tau_*^3/\tau_{\scriptscriptstyle \Omega}^2$ that leads to logarithmic multiplier: $ {a}_{\rho\rho} \sim \tau_{\scriptscriptstyle\Omega}^2 \ln(\tau_*^3/\tau_{\scriptscriptstyle\Omega}^3)$.

\textit{Now we calculate ${a}_{\varphi\varphi}$.} In both limits integral (\ref{xi_eta_start}) is determined by $q_{\rho, \varphi} \sim 1$ thus we can rewrite the integral:
\begin{eqnarray}\label{detailed_tau_beta_xixi}
    {a}_{\varphi\varphi} \sim \tau_{\scriptscriptstyle\Omega}^{-3}
    \itl^{\infty}_{\sim 1} d\tau \tau^2
    \itl^1_{\sim\mathrm{Ro}}\frac{dq_z}{q_z^2 +(\tau/\tau_{\scriptscriptstyle\Omega})^6} e^{-\Gamma}
    =
    \itl^{\infty}_{\sim 1}  \frac{d\tau \tau^2}{\tau^3} \left( \mathop{\mathrm{arctg}}\left(\frac{\tau_{\scriptscriptstyle\Omega}^3}{\tau^3}\right) - \mathop{\mathrm{arctg}}\left(\frac{\tau_{\scriptscriptstyle\Sigma}^3}{\tau^3}\right)\right) \mathcal{G}\left(\frac{\tau_*}{\tau} \right)
\end{eqnarray}
where $\mathcal{G}$ is of the order of unity at $\tau\lesssim\tau_*$ and rapidly decreases at $\tau\gtrsim\tau_*$.

In the limit $\mathrm{Ro}\ll \mathrm{Pr}_{\mathrm{m}}$. the integral is accumulated at $q_z \sim \left(\tau/\tau_{\scriptscriptstyle\Omega}\right)^3$ and $\mathop{\mathrm{max}}\left\{ 1, \tau_{\scriptscriptstyle\Sigma} \right\}\lesssim\tau\lesssim\tau_*$  correspondingly. The integrand depends on time as $\sim 1/\tau$ at $\mathop{\mathrm{max}}\left\{ 1, \tau_{\scriptscriptstyle\Sigma} \right\}\lesssim\tau\lesssim \tau_*$ and decreases faster due to the viscous exponent at $\tau\gtrsim \tau_*$ after the integration over all components of wave-vector. Note that $\varphi\varphi-$element's integrand depends on time as $\sim 1/\tau$ that leads to logarithmic behaviour:
\begin{eqnarray}\label{eq:103}
    {a}_{\varphi\varphi}
    \sim
    \ln\left(\tau_*/\mathop{\mathrm{max}}\left\{1, \tau_{\scriptscriptstyle\Sigma}\right\}\right)\sim\frac{1}{3}\ln\left(\mathop{\mathrm{min}}\left\{ \mathrm{Ro}/\mathrm{Ek}
    ,
    \mathrm{Pr}_{\mathrm{m}}^{-1} \right\}\right)
\end{eqnarray}

In the opposite limit $\mathrm{Ro}\gg \mathrm{Pr}_{\mathrm{m}}$, $\varphi\varphi$-element is determined by $q_z\sim (\tau/\tau_{\scriptscriptstyle\Omega})^3$, $\mathop{\mathrm{max}}\left\{1,\tau_{\scriptscriptstyle\Sigma}\right\}\lesssim \tau \lesssim \tau_{\scriptscriptstyle\Omega}$. After the integration over wave-vectors integrand in (\ref{eq:103}) depends on time as $\sim 1/\tau$ at $\mathop{\mathrm{max}}\left\{1,\tau_{\scriptscriptstyle\Sigma} \right\}\lesssim \tau \lesssim \tau_{\scriptscriptstyle\Omega}$, as $\sim \tau_{\scriptscriptstyle\Omega}^3/\tau^4$ at $\tau_{\scriptscriptstyle\Omega} \lesssim \tau \lesssim \tau_*$ and decreases faster at larger times due to the viscous exponent. Therefore in the limit $\mathrm{Ro}\gg \mathrm{Pr}_{\mathrm{m}}$ (i.e. $\tau_*\gg\tau_{\scriptscriptstyle\Omega}$) $\varphi\varphi-$ element is determined by times $\mathop{\mathrm{max}}\left\{1,\tau_{\scriptscriptstyle\Sigma} \right\} \lesssim \tau \lesssim \tau_{\scriptscriptstyle\Omega}$:
\begin{eqnarray}\label{beta02}
    {a}_{\varphi\varphi} \sim
    \ln\left(\tau_{\scriptscriptstyle\Omega}/\mathop{\mathrm{max}}
    \left\{1, \tau_{\scriptscriptstyle\Sigma}\right\}\right) \sim \frac{1}{3}\ln\left(\mathop{\mathrm{min}}\left\{\mathrm{Pr}_{\mathrm{m}}/\mathrm{Ek}, \mathrm{Ro}^{-1}\right\}\right)
\end{eqnarray}

\paragraph{Account for finite size of the vortex.} Let us take into account the finite passage time $\tau_t$ for inertial waves, see the first paragraph in Section~\ref{sec:discussion}. At times $\tau \gtrsim \tau_{t}$ developed theory \cite{kolokolov2020structure, ogorodnikov2022structure} does not work and $\tau_{t}$ becomes the maximum possible upper limit in all time integrals. Further we consider only $\varphi\varphi$-elements, see (\ref{detailed_tau_beta_xixi}), that is relevant for the large-scale dynamo. Expression (\ref{detailed_tau_beta_xixi}) depends on only $\mathop{\mathrm{min}}\left\{ \tau_{\scriptscriptstyle\Omega}, \tau_* \right\}$. Thus it is influenced by finite $\tau_t$ only if $\tau_{t} \lesssim \mathop{\mathrm{min}}\left\{ \tau_{\scriptscriptstyle\Omega}, \tau_* \right\}$, otherwise ${a}_{\varphi\varphi}$ remains unchanged. If $\tau_t\ll 1$, then the large-scale vortex flow is not relevant and the wave statistics is axially symmetric. So below we consider $\tau_t\gg1$ case.

If $\tau_t>\tau_{\scriptscriptstyle \Sigma}$, then integral (\ref{detailed_tau_beta_xixi}) has logarithmic behavior and is determined by times $\mathop{\mathrm{max}}\left\{1, \tau_{\scriptscriptstyle\Sigma}\right\} \lesssim \tau \lesssim \tau_{t}$. Thus ${a}_{\varphi\varphi} \sim \ln \left(\tau_{t}/ \mathop{\mathrm{max}}\left\{1, \tau_{\scriptscriptstyle\Sigma}\right\}\right)$. If $1 \ll \tau_{t} \ll \tau_{\scriptscriptstyle\Sigma}$, the arguments of $\mathop{\mathrm{arctg}}$ in the expression (\ref{detailed_tau_beta_xixi}) are large therefore the expression in brackets can be estimated as $\sim \tau^3/ \tau_{\scriptscriptstyle\Sigma}^3$. The time-dependent part of the integral is $\sim \tau^2$, and ${a}_{\varphi\varphi}$ depends on upper time limit $\tau_{t}$ according to the power law, ${a}_{\varphi\varphi} \sim \left(\tau_{t}/\tau_{\scriptscriptstyle\Sigma}\right)^3$.

\paragraph{Calculation of $\rho\varphi$-element.} Integrand in (\ref{beta_start_general_0}) is sign-variable function. Let's first take integral over $\tau$ by parts and then take it over $\bm{q}$.
\begin{eqnarray}\label{rhophiint}
    a_{\rho\varphi} = \tau_{\scriptscriptstyle \Omega}^{-3} \itl (d^3q) q\chi(q) \itl^\infty_0 d\tau \frac{q_\varphi q^\prime_\rho q^{\prime 2}}{\left(q^\prime/\tau_{\scriptscriptstyle\Omega}\right)^6+q_z^2}e^{-\Gamma}
    &=&
    \frac{\varsigma \tau_{\scriptscriptstyle \Omega}^{-3}}{2} \itl (d^3q) q\chi(q) \itl_{q^2}^{\infty} \frac{d(q^{\prime 2})q^{\prime 2}}{ (q^{\prime }/\tau_{\scriptscriptstyle\Omega})^6 + q_z^2}e^{-\Gamma}
\end{eqnarray}
If $\mathrm{sign}\left(\varsigma q_\rho q_\varphi\right)>0$, variable $q^{\prime 2}$ growth monotonically with time. If $\mathrm{sign}\left(\varsigma q_\rho q_\varphi\right)<0$, variable $q^{\prime 2}$ monotonically decreases at $\tau \in \left(0, -\varsigma q_\rho/q_\varphi \right)$ from $q^2$ to $(q_z^2+q_\varphi^2)$ (both are of the order of unity) and then it increases monotonically at $\tau > -\varsigma q_\rho/q_\varphi$ from $(q_z^2+q_\varphi^2)$ to infinity. Integral (\ref{rhophiint}) is essential in the region where the both terms in denominator are of the same order, so $q^{\prime 3} \sim \tau_{\scriptscriptstyle \Omega}^3 q_z$. As we consider limit $\tau_{\scriptscriptstyle \Omega}\gg 1$, integral is accumulated at large $q^\prime$ that corresponds to $\tau\gg 1$. So, times $\tau \gg 1$ are relevant thus we can put $q^{\prime 2}\approx q^2_\varphi \tau^2$ and get
\begin{eqnarray}\label{rhophiint2}
    a_{\rho\varphi} \sim \varsigma \tau_{\scriptscriptstyle \Omega}^{-3} \itl (dq_\rho) \itl (dq_\varphi) \itl^\infty_1 d\tau (q_\varphi \tau)^4 \itl \frac{dq_z}{q_z^2 + (q_\varphi\tau/\tau_{\scriptscriptstyle \Omega})^6} q\chi(q) e^{-\Gamma}.
\end{eqnarray}
Similarly to $a_{\rho\rho}$ and $a_{\varphi\varphi}$, integral over $q_z$ in $a_{\rho\varphi}$ is determined by $q_z \sim (q_\varphi \tau/\tau_{\scriptscriptstyle \Omega})^3$ and equals to $\sim (q_\varphi \tau/\tau_{\scriptscriptstyle \Omega})^{-3}$. Integrals over $q_{\rho,\varphi}$ are determined by $q_{\rho,\varphi}\sim 1$.

In the limit $\mathrm{Ro}\ll\mathrm{Pr}_\mathrm{m}$ time-dependent part of integrand is $\sim \tau$ at $\tau\lesssim \tau_*$ and then rapidly decreases at $\tau\gtrsim \tau_*$ due to viscous exponent. So integral (\ref{rhophiint2}) is determined by times $\tau\sim \tau_*$ that corresponds to $q_z \sim (\tau_*/\tau_{\scriptscriptstyle \Omega})^3 \sim \mathrm{Ro}/\mathrm{Pr}_\mathrm{m}\gg \mathrm{Ro}$ in three-dimensional sector of wave-vector space. Thus in this limit we have $a_{\rho\varphi} \sim \varsigma\tau_*$.

In the limit $\mathrm{Ro}\gg\mathrm{Pr}_\mathrm{m}$ time-dependent part of integrand is $\sim \tau$ at $\tau\lesssim \tau_{\scriptscriptstyle \Omega}$ and then decays at $\tau\gtrsim \tau_{\scriptscriptstyle \Omega}$ due to the influence of $\chi(q)$ that rapidly decreases at $q_z \gtrsim 1$. Therefore integral (\ref{rhophiint2}) is determined by times $\tau\sim \tau_{\scriptscriptstyle \Omega}$ that corresponds to $q_z \sim 1$. So, in this limit off-diagonal element is of the order of $a_{\rho\varphi} \sim \varsigma\tau_{\scriptscriptstyle \Omega}$.

\subsection{Case $\mathrm{Pr}_\mathrm{m}\ll\mathrm{Ek}$}

This case corresponds to $\tau_{\scriptscriptstyle\Omega}\ll 1$. Let's first consider diagonal elements (\ref{beta_start_general_0}). Its integrand is positive at the whole integration area. As $\tau_{\scriptscriptstyle\Omega}\ll 1$ and $q_z^2 \leq q^{\prime 2}$, the term $q_z^2$ in denominator can be neglected. So pre-exponent of the integrand depends on time as $\sim q^{\prime}_{i} q^{\prime}_{j}/q^{\prime 4}$:
\begin{eqnarray}
    {a}_{ij} \sim \tau_{\scriptscriptstyle\Omega}^3 \itl^\infty_0 d\tau \itl (d^3q)  \frac{q\chi(q) q^{\prime}_i q^{\prime}_j}{q^{\prime 4}} e^{-\Gamma}
\end{eqnarray}
For all diagonal elements integral over wave-vectors is accumulated at $q_{\rho,z }\sim 1$.
Let's calculate zz-element first. The main contribution to the integral is accumulated at $q_\varphi\sim 1/\tau$. Thus time-dependent integrand has form $\sim 1/\tau$ at times $1\lesssim \tau \lesssim \tau_*^3$ . The upper limit of this time range is determined by pre-exponent and corresponds to \textquotedblleft viscous times\textquotedblright \cite{ogorodnikov2022structure}. It leads to the emergence of a logarithmic multiplier in ${a}_{zz}$. Similar results can be obtained for ${a}_{\rho\rho}$: it is also accumulated at $q_\varphi\sim 1/\tau$ and $1\lesssim \tau \lesssim \tau_*^3$ and it equals to ${a}_{\rho\rho} \sim {a}_{zz} \sim \tau_{\scriptscriptstyle\Omega}^3 \ln\tau_*$.

The main contribution to ${a}_{\varphi\varphi}$ is accumulated at $q_\varphi\sim 1$ and $\tau\sim 1$. {Note also that in this limit timescale $\tau_{t} \gtrsim 1$ that corresponds to finite length of inertial waves has no influence on the expression of ${a}_{\varphi\varphi}$ because this element is determined by $\tau\sim 1 \lesssim \tau_{t}$. So ${a}_{\varphi\varphi}$ does not depend on $\tau_{t}$ and remains the same as in the limit $\tau_{t} \to \infty$.}  Therefore ${a}_{\varphi\varphi}\sim \tau_{\scriptscriptstyle\Omega}^3$.

Off-diagonal element ${a}_{\rho\varphi}$ has following form:
\begin{eqnarray}
    {a}_{\rho\varphi} \sim \tau_{\scriptscriptstyle\Omega}^3 \itl (d^3q) q\chi(q) \itl^\infty_0 d\tau \frac{ q_\varphi q_\rho^\prime}{q^{\prime 4}} e^{-\Gamma}
\end{eqnarray}
It has sign-variable function in the integrand. Therefore it is convenient to take integral over $\tau$ by parts:
\begin{eqnarray}
    \itl^\infty_0 d\tau \frac{ q_\varphi q_\rho^\prime}{q^{\prime 4}} e^{-\Gamma} = \frac{\varsigma}{2} \itl \frac{dq^{\prime 2}}{q^{\prime 4}} e^{-\Gamma} = \frac{\varsigma}{2}\left\{ \frac{1}{q^2} - \tau_*^{-3}\itl^\infty_0 d\tau e^{-\Gamma}
    \right\}
\end{eqnarray}
This integral is accumulated at $\tau\sim 1$, so
\begin{eqnarray}
    {a}_{\rho\varphi} \sim \varsigma\tau^3_\Omega \itl (d^3q) \frac{\chi(q)}{q} \sim \varsigma\tau^3_\Omega
\end{eqnarray}
This element is determined by $q_{\varphi, \rho, z} \sim 1$.

Therefore in the limit $\mathrm{Pr}_\mathrm{m}\ll\mathrm{Ek}$ off-diagonal element is ${a}_{\rho\varphi} \sim -\varsigma\tau_{\scriptscriptstyle\Omega}^3 $.


\end{document}